\documentstyle[11pt,aaspp4]{article}

\newcommand{\etal}{{\it et al.\/}}
\begin{document}

\title{Star Formation in the Hubble Deep Field North}

\author{Judith Cohen\altaffilmark{1}}

\altaffiltext{1}{California Institute of Technology, 
Pasadena, California, USA}

\begin{abstract} I am currently analyzing the emission line spectra
of the $\sim$600 galaxies from the sample of Cohen \etal\ (2000) and
Cohen (2001) in the region of the HDF-North with $z<1.5$.  
A progress report on this 
effort of the Caltech Faint Galaxy Redshift Survey is presented.

\end{abstract}

\section{For Proceedings of meeting: Galaxy Evolution: Theory
and Observations, Cozumel, March 2002}

I am measuring the equivalent widths of emission lines in the $\sim$600
galaxies from the sample of Cohen \etal\ (2000) and
Cohen (2001) in the region of the HDF-North with
$z<1.5$. This redshift survey,
based on spectra from LRIS at the Keck
Observatory, is more than 93\% complete
for $R < 24$ in the HDF itself and for $R < 23.5$ in the Flanking
Fields, comprising a circle 8 arcmin in diameter centered on the HDF.
While all the measurements are carried out automatically with a script, they
must be checked by hand.  So far I've completed 
checking the results for 254 galaxies, all with spectra taken by me.

The analysis of these emission line strengths is in progress.  
Equivalent widths have been measured both using a feature
and blue and red continua bandpasses defined in the rest frame 
as well as using Gaussian fitting.

We adopt the emitted luminosity in the 3727 \AA\ line of [OII]
as our primary indicator of the star formation rate
because of the wider redshift range over which it can be detected,
given that  at present we are restricted only to optical spectra for galaxies
in our sample.

All luminosities are given in the rest frame, unless otherwise specified.
The SED formalism of Cohen (2001) is used to derive all
continuum luminosities.
Emission line luminosities are calculated using the 
measured equivalent width
and the inferred continuum luminosity deduced from broad band large
aperture photometry. We are thus assuming that the 1 arc-sec wide
slit passes a reasonable sample of the total galaxy light, i.e. that emission
is not concentrated just in the region of the galactic nucleus.

As an example of the analyses that can be carried out
with this data, Figure 1 shows
$L(3727)$ (luminosity of the [OII] emission line in the rest frame) 
as a function of redshift.
An arrow indicates the luminosity in this emission line 
corresponding to a 
SFR of 100 $M${\mbox{$_{\odot}$}}/yr.  The mean local star formation
rate for emission line galaxies from the Universidad Complutense de
Madrid Survey (Gil de Paz \etal\ 2000) is indicated as well.  These 
local SFR measures
include reddening corrections, while our data do not.
The effect of a reddening correction corresponding to the mean adopted
$E(B-V)$ of 0.5 mag is indicated on the figure.  Adopting this value
as a typical reddening, we find SFR rates among the galaxies
in this distant sample which cover approximately the same range
as in local samples of emission line galaxies.

In Figure 2 we show
the star formation rate per unit mass (SFR/{\it{M}}) for these 254 galaxies.
This  parameter is
an indicator of the current SFR divided by the mean SFR over the lifetime
of the galaxy.  We use the rest frame luminosity emitted in the 3727 \AA\
[OII] emission line divided by the rest frame luminosity at $K$ 
as a measure of the specific SFR.  The former is dominated by
emission from the youngest stars in a galaxy, with high luminosity
per unit mass, while the latter
is dominated by the light from any older population that may be present,
and is much less affected by the possible presence of luminous young stars.

The specific star formation rate parameter is plotted
as a function of redshift.  As suggested by Guzman \etal\ (1997)
based on a much smaller sample of galaxies, the
upper bound of the specific SFR is independent of redshift. 

However, the luminosity of those galaxies with the highest specific SFR
increases with redshift.  At the present epoch, high specific SFR is
associated primarily with low luminosity (presumably low mass) galaxies,
but at $z \sim 1$, the most luminous galaxies show this high specific SFR.

To summarize the results we have found thus far, our data 
in the HDF demonstrate that 
star formation in distant galaxies is quite similar to that in
the local Universe.  The range of equivalent widths of the
most common diagnostic lines is similar over the full
range of redshift probed here, $z<1.3$.  
We define the specific SFR as the
star formation rate per unit luminosity (equivalent to
SFR per unit mass for a constant ${\it{M}}/L$ ratio, a reasonable
approximation for luminosity at rest frame $K$).  This ratio is 
effectively the current SFR divided by the mean
SFR over the lifetime of the galaxy.
The maximum specific SFR over the full range of redshift
probed here is similar to that
of local star forming galaxies.

However, the most massive galaxies seen at $z\sim1$ are 
forming stars at this rate while locally, high 
specific star formation rates are
seen mostly in low mass galaxies.
It appears that the high specific SFR
seen in galaxies at $z\sim1$ is associated with
a mode of star formation in which the whole galaxy, not
just the nucleus or a few isolated HII regions, participates.

The fraction of narrow-lined AGNs is small ($\le15$\%
of the sample), out to $z<0.4$.
There is no evidence from the present sample
to support a larger fraction for $z>0.4$.

I hope to complete the check of emission line strengths
for the full set of spectra within a few months.

\clearpage

\begin{figure}
\epsscale{0.9}
\plotone{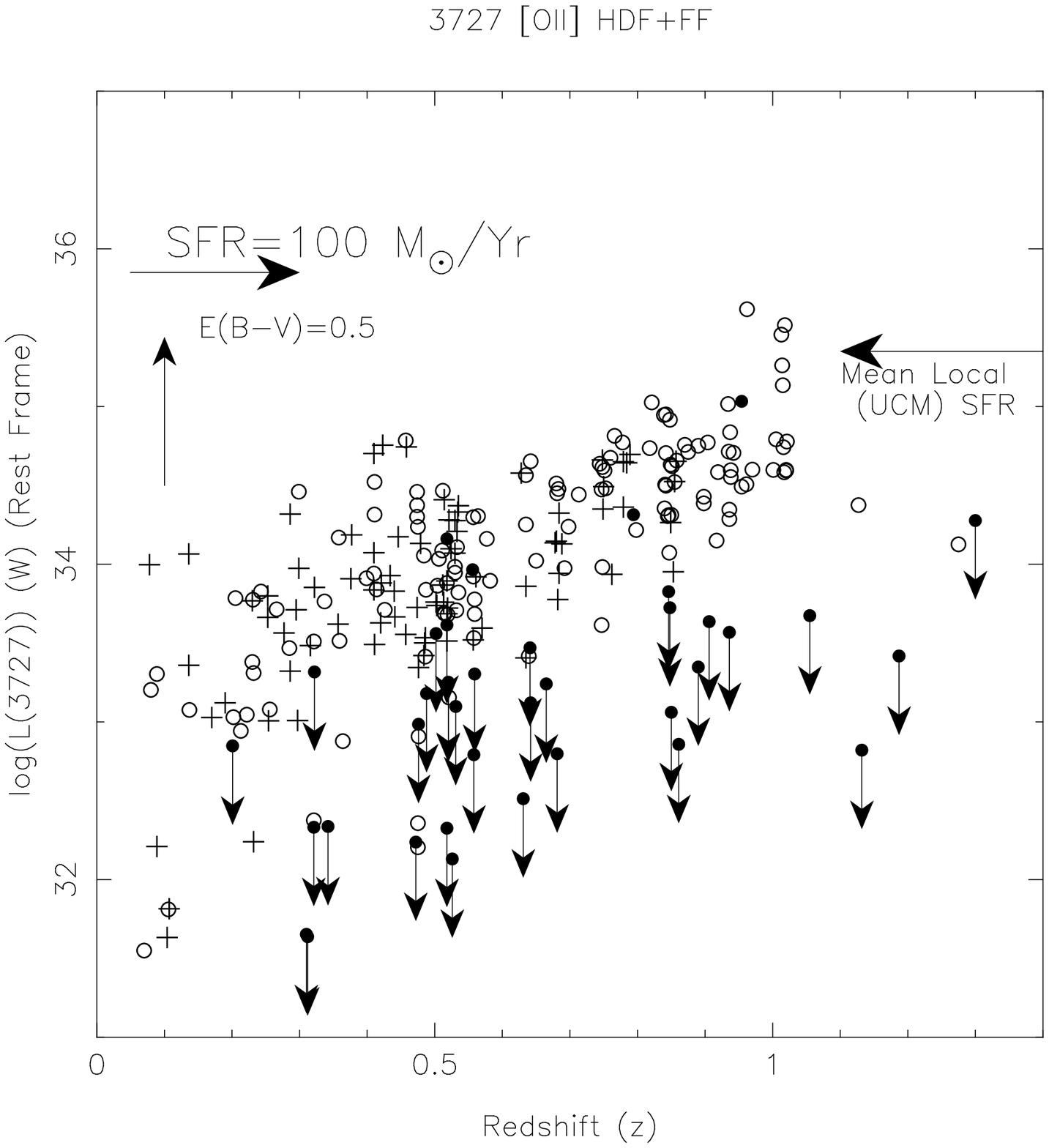}
\caption[]{The emitted flux in the rest frame in the 3727 [OII]
line is shown as a function of redshift.  Open circles are galaxies
whose spectra are dominated by emission lines, filled circles are
galaxies lacking emission lines (with the 3727 \AA\ [OII] emission
line flux shown as upper limits),
while crosses denote intermediate cases.
\label{fig_lum3727}}
\end{figure}

\begin{figure}
\epsscale{0.9}
\plotone{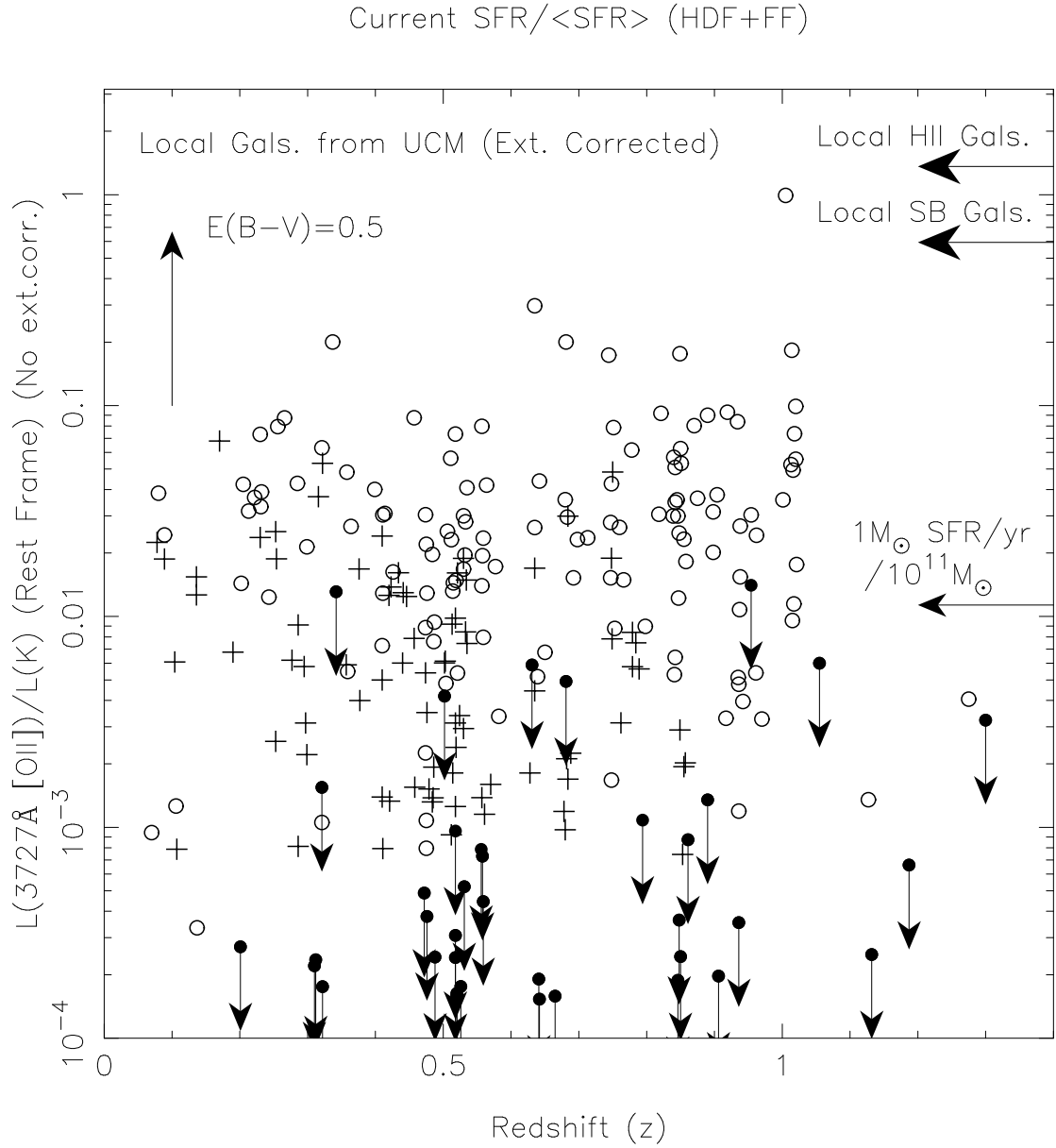}
\caption[]{The star formation rate per unit stellar mass for
$\sim$250 galaxies in the HDF-N is shown as a function of redshift.
The symbols are as in Figure 1.
\label{fig:sfr_per_mass} }
\end{figure}

\end{document}